\documentclass[conference,a4paper]{IEEEtran}
\IEEEoverridecommandlockouts
\usepackage{cite}
\usepackage{amsmath,amssymb,amsfonts}
\usepackage{algorithmic}
\usepackage{graphicx}
\usepackage{textcomp}
\usepackage{xcolor}
\usepackage{upgreek}
\usepackage{verbatim}
\usepackage[absolute]{textpos}
\TPMargin{5pt}

\def\BibTeX{{\rm B\kern-.05em{\sc i\kern-.025em b}\kern-.08em
    T\kern-.1667em\lower.7ex\hbox{E}\kern-.125emX}}
\DeclareMathOperator{\EV}{\mathbb{E}}    
\DeclareMathOperator{\Real}{\mathbb{R}}    

\DeclareMathOperator{\valpha}{\boldsymbol{\alpha}}  
  
\DeclareMathOperator{\vpi}{\boldsymbol{\pi}}  
\DeclareMathOperator{\vmu}{\boldsymbol{\mu}}  
  
\DeclareMathOperator{\vdelta}{\boldsymbol{\delta}}    
\DeclareMathOperator{\vy}{\boldsymbol{y}}    
    
\DeclareMathOperator{\vX}{\boldsymbol{X}}    
    
\DeclareMathOperator{\vb}{\boldsymbol{b}}    
 
\DeclareMathOperator{\vg}{\boldsymbol{g}} 
\DeclareMathOperator{\vs}{\boldsymbol{s}}   
\DeclareMathOperator{\vv}{\boldsymbol{v}}    
    
\DeclareMathOperator{\vV}{\boldsymbol{V}}

\DeclareMathOperator{\vd}{\boldsymbol{d}}   
\DeclareMathOperator{\vu}{\boldsymbol{u}}   
   
\DeclareMathOperator{\setT}{\mathcal{T}}    

\DeclareMathOperator{\setS}{\mathcal{S}}

\DeclareMathOperator{\ftint}{\forall t \in \setT}
\newcommand{\EVof}[1]{{\EV  \left[ #1  \right]}}

\newcommand{\copyrightstatement}{
    \begin{textblock}{0.84}(0.08,0.93)    
         \noindent
         \footnotesize
© 2022 IEEE.  Personal use of this material is permitted.  Permission from IEEE must be obtained for all other uses, in any current or future media, including reprinting/republishing this material for advertising or promotional purposes, creating new collective works, for resale or redistribution to servers or lists, or reuse of any copyrighted component of this work in other works.  \end{textblock}
}    

\begin{document}
\copyrightstatement

\title{
Bidding and Scheduling in Energy Markets: \\
Which Probabilistic Forecast Do We Need?
\thanks{This work is funded by the German federal Ministry for Economic Affairs and Energy. Funding code: 03ET1638.}
}

\author{\IEEEauthorblockN{Mario Beykirch, Tim Janke and Florian Steinke}
\IEEEauthorblockA{\textit{Energy Information Networks and Systems
} \\
\textit{TU Darmstadt}\\
Darmstadt, Germany \\
\{mario.beykirch, tim.janke, florian.steinke\}@eins.tu-darmstadt.de}
}

\maketitle

\begin{abstract}
Probabilistic forecasting in combination with stochastic programming is a key tool for handling the growing uncertainties in future energy systems.
Derived from a general stochastic programming formulation for the optimal scheduling and bidding in energy markets we examine several common special instances containing uncertain loads, energy prices, and variable renewable energies. 
We analyze for each setup whether only an expected value forecast, marginal or bivariate predictive distributions, or the full joint predictive distribution is required.
For market schedule optimization, we find that expected price forecasts are sufficient in almost all cases, while the marginal distributions of renewable energy production and demand are often required.
For bidding curve optimization, pairwise or full joint distributions are necessary except for specific cases. 
This work helps practitioners choose the simplest type of forecast that can still achieve the best theoretically possible result for their problem and researchers to focus on the most relevant instances. 
\end{abstract}

\begin{IEEEkeywords}
stochastic programming, probabilistic forecasting, energy forecasting, energy scheduling 
\end{IEEEkeywords}

\section{Introduction}
\begin{figure}[t]
    \centering
    \includegraphics[width=0.8\linewidth]{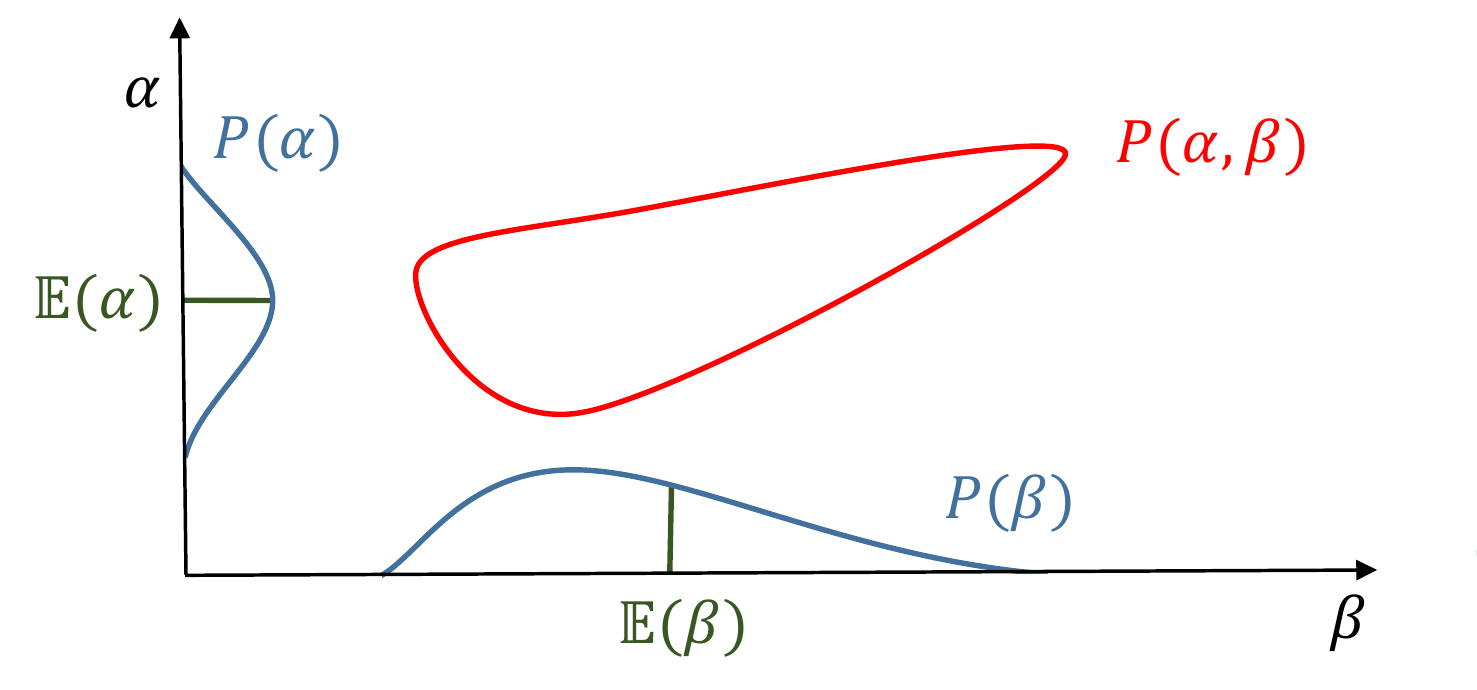}
    \caption{Forecasts can cover the expected values (green), the marginal distributions (blue), or the full joint (red) distribution of the considered uncertain factors like demands, market prices, or renewable energy production.
    }
    \label{fig:intro_distExamples}
\end{figure}

Stochastic programming based on probabilistic forecasts is often seen as a key tool for the optimal marketing and operation scheduling of energy systems under uncertainty, see e.g. \cite{Maggioni2012_analyzingEvSolution,kumbartzky2017_OptimalChpOperation,Ottesen2016_prosumerBidding,Ackermann2019}.
Probabilistic forecasts improve over classic point forecasts, which usually only report the expected value, by predicting a probability distribution to account for the associated forecast uncertainty \cite{gneiting2014_probabilisticFc}.
This distribution can either be a univariate marginal distribution corresponding to a certain dimension, e.g., in time or space, or take the form of a multivariate joint distribution, see Figure~\ref{fig:intro_distExamples}.
When employing these approaches,
one might feel inclined to use the most sophisticated forecasting methods available, e.g. \cite{dumas2022_deepGenModProbFc,toubeau2018_deepLearningProbFc,janke2020_priceFcImplGenMods},
to forecast the full joint distribution of all uncertain factors such as market prices, variable renewable energy production, and demands.
However, using such a complex forecast may often be computationally expensive and difficult to implement. 
We will show theoretically that in many cases simpler forecasts of the expected value or of marginal distributions will lead to equally optimal results.
The efforts for producing the forecast as well as for using it in the optimization process will then be much lower.
The resulting performance will even be superior in practice,
since high-dimensional distributions require a larger amount of samples 
and thereby increase the computational costs of the optimization. 
Hence, it is important to know which type of forecast is actually required for the task at hand. 


The goal of this paper is to examine for various common optimal scheduling and bidding setups in energy markets
which type of forecast is minimally required,
covering the range from the full or pairwise joint distribution, over the  marginal distributions, to the expected value-only forecast.
To this end, we first introduce a general stochastic programming formulation with uncertain demand, renewable energy production, and electricity prices.
We then examine different special cases for market schedule or bidding curve optimization, including power plants with and without start cost, systems with and without storage, and different imbalance cost schemes. 
We furthermore identify key features that indicate which type of forecast is required. 


We find that the full joint forecast distribution is rarely required.
For optimal market scheduling, no probabilistic price prediction is required at all.
Bidding curve optimization almost always requires joint forecast distributions between the price, the demand, and the renewable generation.
A storage necessitates a full temporal joint distribution, a power plant with start up cost only requires distributions that are joint between adjacent time steps.
In a simulation experiment for scheduling a power plant with a varying share of startup costs, we demonstrate that the necessity for probabilistic forecasts changes gradually depending on
the power plant's flexibility. 

Various stochastic programming papers compare their performance against a deterministic model that uses only point forecasts,
e.g., for bidding curve optimization based on full multivariate probabilistic forecasts \cite{Ottesen2016_prosumerBidding,Ackermann2019, kumbartzky2017_OptimalChpOperation}.
\cite{Maggioni2012_analyzingEvSolution} examines whether the solution of the deterministic model offers any value for the stochastic problem.
However, quantitatively comparing the optimal values of the deterministic model and the stochastic model does not explain the underlying reasons for the added value of the stochastic problem.
To our knowledge, this paper is the first to systematically examine which type of forecast is minimally required for different stochastic problem formulations for energy systems  to achieve the best possible results.
Our work thus contributes to the field of research on the valuation of (probabilistic) forecasts \cite{Hong2020_EnergyForecastingOutlook}.

Section \ref{sec:background} introduces notation and background theory.
Section \ref{sec:general} presents the general stochastic programming formulation. 
In sections \ref{sec:schedule} and \ref{sec:bidding} we examine specific setups for schedule  and bidding curve optimization, respectively.
A bidding curve simulation experiment is presented in \ref{sec:experiment}. 
Key results are discussed and summarized in section \ref{sec:discuss}.

\section{Background \& Notation}
\label{sec:background}
A two stage stochastic optimization problem \cite{Shapiro2021_lecturesOnStochProg} can be expressed as
\begin{align}
    \min_{ \vX, \vy} &\quad \EVof{g(\omega,\vX,\vy(\omega))} \\
    \mathrm{s.t.}& \quad \vX \in \mathcal{X}, \; \vy(\omega) \in \mathcal{G}(\omega,\vX) \; \mathrm{a.e.} \; \omega \in \Omega ,  
\end{align}
with the first stage decision variables $\vX$, the second stage decision variables $\vy(\omega)$, and the cost function $g(\omega,\vX,\vy(\omega))$. 
$\mathcal{X}$ comprises all constraints on $\vX$ and $\mathcal{G}(\omega,\vX)$ comprises all constraints on $\vy(\omega)$ given the first stage decision $\vX$ and a realization of the random vector $\valpha(\omega)$.
The random vector $\valpha: \Omega \rightarrow \Real^{k}$ with dimension $k$ is defined on a probability space $(\Omega,\mathcal{F},P)$, with sample space $\Omega$, $\sigma$-algebra $\mathcal{F}$, and probability measure $P$. 
For conciseness we will write $\valpha(\omega)$ as $\valpha^\omega$.

For the random vector $\valpha^\omega$ we denote 
the (forecasted) full joint distribution by $P(\valpha^\omega)$, the pairwise joint distributions by $\{P(\alpha_i^\omega,\alpha_{j}^\omega)\}_{i,j=0}^{k}$, the marginal distributions by $\{P(\alpha_i^\omega)\}_{i=0}^{k}$, the expected value of each component by $\{\EVof{\alpha_i^\omega}\}_{i=0}^{k}$.
Note that a set of pairwise joint distributions must have consistent univariate marginal distributions, i.e., if a random variable is described by more than one multivariate distribution, all marginal distributions with regard to this random variable must be equal.

The expected value of (a function of) random variable $\valpha^\omega$ is
\begin{equation}
\EVof{\vu(\valpha^\omega)}=\int_{\Omega} \vu(\valpha^\omega) dP_\omega.
\end{equation}
The expectation is linear, i.e., $\EVof{c_1 \vu(\valpha^\omega) + c_2 \vv(\valpha^\omega)}= c_1 \EVof{\vu(\valpha^\omega)} + c_2 \EVof{\vv(\valpha^\omega)}$ for all maps $\vu(.)$ and $\vv(.)$ and constants $c_1$ and $c_2$.
If two random variables $\alpha_i^\omega$ and $\alpha_j^\omega$ are independent, $\EVof{\alpha_i^\omega \alpha_j^\omega} =\EVof{\alpha_i^\omega} \EVof{\alpha_j^\omega}$ holds. 

We define the positive and the negative function as $[x]_+ = \mathrm{max}(x,0)$ and $[x]_- = \mathrm{max}(-x,0)$, respectively.

\section{General Problem Formulation}
\label{sec:general}
\begin{figure}[t]
    \centering
    \includegraphics[width=1\linewidth]{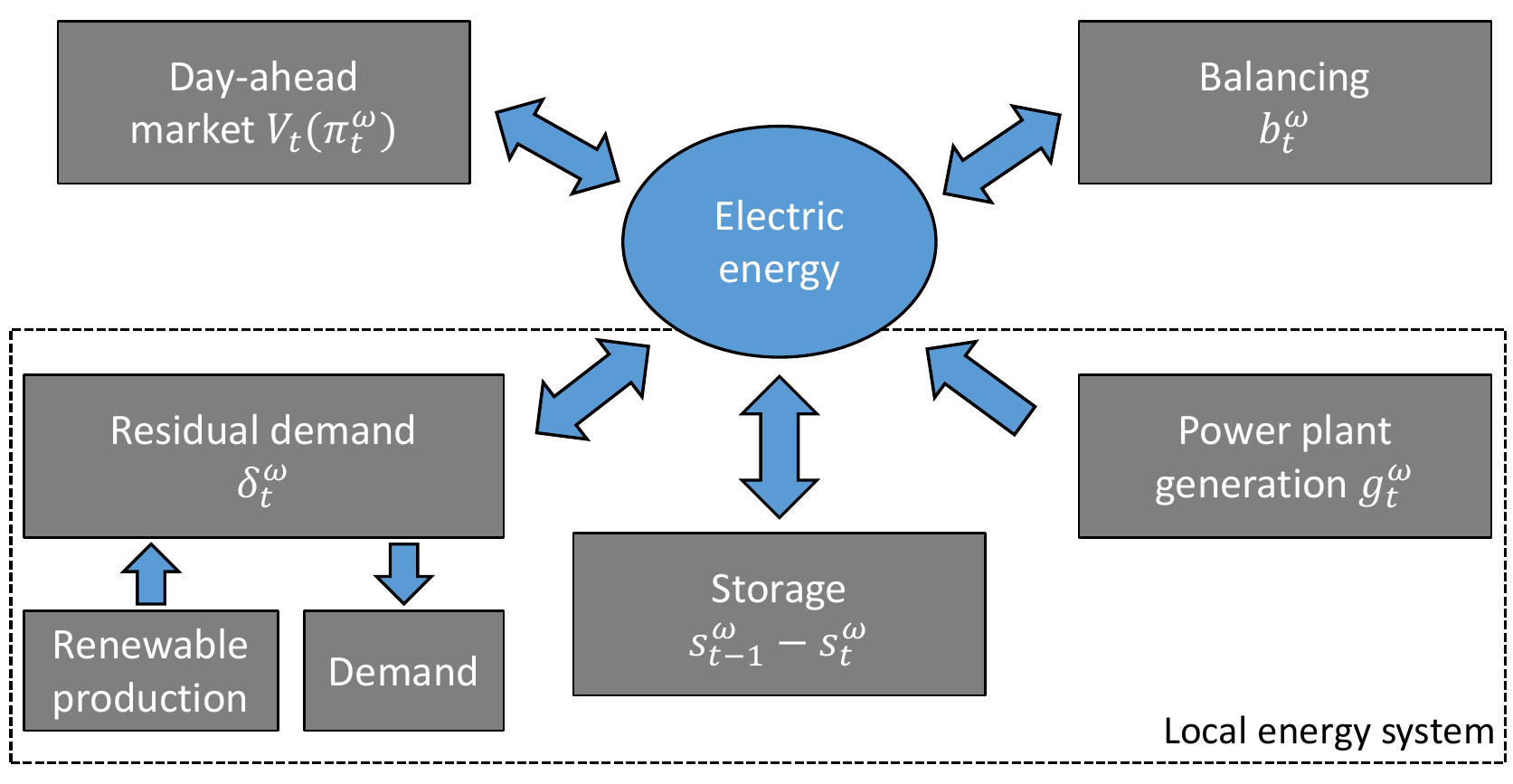}
    \caption{Structure of the considered general problem setup.}
    \label{fig:setup}
\end{figure}

We first introduce a general stochastic programming formulation for the bidding and scheduling of a local energy system in interaction with energy markets. 
In this context we take the perspective of the operator of the local power system and neglect any grid constraints.

The formulation is based on the physical setup shown in Figure~\ref{fig:setup} which includes
one controllable power plant, volatile renewable generation, electricity storage, and a power demand to be satisfied.
The system can participate in the day-ahead electricity market and obtain balancing power from the grid.
This setup covers many common problem instances which can be specified by setting the capacities of the different components accordingly.

Moreover, by assuming some of the parameters to be certain or uncertain we can consider the task of uncertainty management.
These specific instances and the required forecasts for each situation are examined in detail below.

We consider a mixed-integer linear stochastic two stage problem
where the submission of a market bid is the first stage decision and the operation of the system components and the settling of imbalances are the second stage decisions,
\begin{align}
& \begin{aligned}
\min_{\vg^\omega,\vu^\omega,\vs^\omega,\vb^\omega,\vV(.)} \EV  &  \left[ \sum_{t\in  \setT} \pi_{t}^\omega V_{t}(\pi_{t}^\omega)+ c_t^{b+} [b_t^\omega]_+  +  c_t^{b-}  [b_t^\omega]_-  \right.  \\ 
&+ \left. c^F_{t} g^\omega_{t}+c^{ST}_{t} I^{st}(u_{t}^\omega,u_{t-1}^\omega)) \vphantom{ \sum_{t\in  \setT}}  \right], \label{eq:sched_gnrlCostF} 
\end{aligned} \\
& \begin{aligned}
\mathrm{s.t.} \quad &  g_{t}^\omega +V_{t}(\pi_{t}^\omega) 
+s_{t-1}^\omega-s_t^\omega
-\delta_{t}^\omega
 = b_t^\omega,  & &\\
& u_{t}^\omega \underline{g}_{t} \le g_{t}^\omega \le u_{t}^\omega \overline{g}_{t},\quad & &  \\
& 0 \le s_t^\omega \le \overline{s}, \quad  & \\
& V_t(.) \in \mathcal{V}, \; u_{t}^\omega \in \{0,1\} 
\quad   \quad
\forall t \in \setT,&\mathrm{a.e.}\; \omega \in \Omega. \label{eq_stpr_} &\quad\quad\quad
\end{aligned}
\end{align}
The random variables in the stochastic optimization problem are the market price $\pi_{t}^\omega$ and the residual demand $\delta_{t}^\omega$, i.e., the demand minus renewable energy generation.
Note that by this definition, $\delta_t^\omega$ can take positive and negative values.
The first stage optimization variables are (the parameters of) the bidding curve $V_{t}(\pi_{t}^\omega)$.
A bidding curve $V_t$ is an increasing function that returns the volume purchased at the day-ahead market at a realized price $\pi_t^\omega$. $V_t$ is negative if energy is sold. 
The constraints for a valid bid curve are expressed by $V_t(.) \in \mathcal{V}$.
We do not consider block bids.
The second stage variables are the power plant's power generation $g_t^\omega$ and on/off state $u_t^\omega$, the energy storage's state of charge $s_t^\omega$, and the resulting power imbalance $b_t^\omega$ that will be covered by reserve power mechanisms.
Parameters considered as fixed are the power plant's fuel and startup costs $c^{F}_t$ and $c^{ST}_t$, the cost of positive and negative imbalances $c_t^{b+}$ and $c_t^{b-}$, the upper and lower plant minimum and maximum operating limits $\underline{g}_{t} $ and $\overline{g}_{t}$, as well as the storage capacity limit $\overline{s}$.
Grid fees, which are not considered, would be added only to the purchase price, hence purchases and sales would need to be considered as two separate variables.  
The function $I^{ST}(u_{t}^\omega,u_{t-1}^\omega)$ takes the value $1$ if a power plant start occurs at time $t$, as indicated by $u_{t}^\omega$ and $u_{t-1}^\omega$, and $0$ otherwise.  
The set $\setT$ includes all considered time steps $t$ in the planning optimization horizon.
Such stochastic programs are typically solved using the sample average approximation (SAA).
Since SAA is not important for the arguments in this work, we refer the reader to \cite{Shapiro2021_lecturesOnStochProg}.

\section{Market schedule optimization} 
\label{sec:schedule}
We first consider a case in which the operator needs to schedule market sales or purchases day ahead, without a price-dependent bidding curve.
This setup might apply, for example, if 
the local energy system participates in the market via an aggregating middleman that charges wholesale market prices.

Such scheduling can be modeled within our general formulation by making $V_t$ independent of the realized price $\pi_t^\omega$, i.e., by setting $V_t(\pi_t^\omega) = V_t$.
In a fully liquid market the schedule optimization problem can then be simplified to
\begin{align}
\min_{\vg^\omega,\vs^\omega,\vb^\omega,\vV } \EV  &  \left[ \sum_{t\in  \setT} \pi_{t}^\omega V_{t} + c_t^{b+} [b_t^\omega]_+  +  c_t^{b-}  [b_t^\omega]_-  \right.  \nonumber \\ 
&+ \left. c^F_{t} g_{t}^\omega+c^{ST}_{t} I^{st}(u_{t}^\omega,u_{t-1}^\omega) \vphantom{ \sum_{t\in  \setT}}  \right]
\end{align}
\begin{align}
\mathrm{s.t.} \quad &  g_{t}^\omega +V_{t} 
+s_{t-1}^\omega-s_t^\omega
-\delta_{t}^\omega
 = b_t^\omega  &\nonumber\\
& u_{t}^\omega \underline{g}_{t} \le g_{t}^\omega \le u_{t}^\omega \Bar{g}_{t}\quad &  \nonumber\\
& 0 \le s_t^\omega \le \Bar{s} \quad & \nonumber \\
&u_{t}^\omega \in \{0,1\} \quad \quad  \quad  \quad \quad \quad \; \ftint , \mathrm{a.e.}\, \omega \in \Omega .& 
\end{align}
Let us examine several special instances of this setup.

\subsubsection{Uncertain price, fixed residual demand}
We first consider a case in which only the market price $\pi_t^\omega$ is uncertain, but the residual demand is certain, i.e., $\vdelta^\omega = \vd$. 
In this case, the random variable is only as a multiplier of the first stage variables in the cost function.
The feasibility of all second stage optimization variables is independent of the random factor, as is their objective.
Thus, they can all be chosen as equal.
Due to the linearity of the expectation operator, the optimization problem then simplifies to
\begin{align}
\min_{\vg,\vu,\vs,\vb ,\vV } \sum_{t\in  \setT}&   \EVof{   \pi_{t}^\omega } V_{t} + c_t^{b+} [b_t]_+  +  c_t^{b-}  [b_t]_-  \nonumber \\ 
&+  c^F_{t} g_{t}+c^{ST}_{t} I^{st}(u_{t},u_{t-1}) 
\end{align}
\begin{align}
\mathrm{s.t.} \quad &  g_{t} + V_{t} + s_{t-1}-s_t -  d_{t}= b_t  &  \nonumber \\
& u_{t} \underline{g}_{t} \le g_{t}  \le u_{t}  \Bar{g}_{t}\quad & \nonumber \\
& 0 \le s_t  \le \Bar{s} \quad &  \nonumber \\
&u_{t}  \in \{0,1\} \quad &\ftint .
\end{align}
It only contains the expected value of $\pi_t^\omega$. 
Using the full joint forecast distribution of $\pi_t^\omega$ in a stochastic program could therefore not improve the results compared to using the expected value forecast.
It is noteworthy that this example includes many use cases, such as battery arbitrage or the scheduling of one or more power plants with startup costs.
Even more complex cost factors and operational constraints would not change the argument.
\\
\textbf{Required forecast:} Expected price.

\subsubsection{Uncertain price and residual demand, 
no storage, no startup costs}
Now consider scheduling electricity purchases or sales given an uncertain residual demand and a generation unit that has either neglectable startup costs or is always running. This optimization problem reads as
\begin{align}
\min_{\vg^\omega, \vb^\omega, \vV} \quad  & \sum_{t\in  \setT}   \EVof{ \pi_{t}^\omega  } V_{t} + c^{b+} \EVof{ [b_t^\omega]_+ }+ c^{b-} \EVof{ [b_t^\omega]_- }&\nonumber \\ 
&\quad + c^F_{t} \EVof{g_{t}^\omega} & \\
\mathrm{s.t.} \quad & g_{t}^\omega+ V_{t}- \delta_{t}^\omega = b^\omega_t \quad \nonumber \\
&   \underline{g} \le g_{t}^\omega \le \Bar{g}\quad 
\quad\quad\quad\quad 
\ftint , \mathrm{a.e.}\, \omega \in \Omega .
\end{align}
Substituting $b_t^\omega$ in the cost function yields 
\begin{align}
    \min_{\vg,\vV} \quad  \sum_{t\in  \setT}  & \EVof{ \pi_{t}^\omega  } V_{t} + c^{b+} \EVof{ [g_{t}^\omega+ V_{t}- \delta_{t}^\omega]_+ } \nonumber \\
    & + c^{b-} \EVof{ [g_{t}^\omega+ V_{t}- \delta_{t}^\omega]_- } + c^F_{t} \EVof{g_{t}^\omega} . 
\end{align}
The cost function now includes the expectation of non-linear expressions of random variables, namely $\delta_t^\omega$ in the non-linear functions $[.]_+$ and  $[.]_-$.
To calculate this expression, the distribution of $\delta_t^\omega$ is necessary.
But since the non-linear expressions only contains $\delta_t^\omega$ and constraints on it do not depend on $\pi^\omega_t$, the marginal distributions $P(\delta_t^\omega)$ and the expectation of $\pi_t^\omega$ are sufficient.
Using joint distributions of several time steps or of both random variables could not improve the results of the stochastic optimization.
\\
\textbf{Required forecast:} Expected price, marginal residual demand distributions. 
\subsubsection{Uncertain price and residual demand, procurement only, symmetrical imbalance cost}
Depending on the design of the balancing market,  we can examine the following special cases. 
If the imbalance cost is symmetric, i.e., $c^b=c^{b+}=-c^{b-}$, we can use $\EVof{\alpha^\omega} = \EVof{[\alpha^\omega]_+}-\EVof{[\alpha^\omega]_-}$ to simplify the cost function to $\sum_{t\in  \setT}  ( \EVof{ \pi_{t}^\omega  } -  c^{b} ) V_t +c^{b}  \EVof{ \delta_t^\omega }$.
Since the term $c^{b}  \EVof{ \delta_t^\omega }$ is a constant summand in the cost function, the optimal schedule is independent of it and a prediction for $\delta_t^\omega$ is therefore not necessary. 
Note that this would result in arbitrage between the day-ahead market and imbalance market, which is not a realistic use case. 
In reality the imbalance cost is often uncertain, such as for the German balancing market. 
The cost function then is
\begin{equation}
    \min_{\vV} \quad  \sum_{t\in  \setT}   (\EVof{ \pi_{t}^\omega} -\EVof{ \beta_t^{\omega}})  V_t + \EVof{ \delta_t^\omega \beta_t^{\omega} },
\end{equation}
with the uncertain imbalance cost $\beta^{\omega}_t$.
Only the expected values of $\delta_t^\omega$ and $\beta_t^\omega$ are in the cost function. 
Hence, the expected value forecast of each is sufficient.\\
\textbf{Required forecast:} Expected price, expected imbalance cost.

\subsubsection{Uncertain price and residual demand, 
no storage, but startup costs}
Next we consider a power plant with startup costs, which results in the stochastic program
\begin{align}
\min_{\vg^\omega,\vu^\omega,\vb^\omega,\vV} \quad& \sum_{t\in  \setT}   \EVof{ \pi_{t}^\omega}  V_{t} + c^{b+}   \EVof{ [b_t^\omega]_+ }+ c^{b-} \EVof{ [b_t^\omega]_- } \nonumber \\
& \quad \quad + c^F_{t}  \EVof{ g_{t}^\omega } + c^{ST}_{t}  \EVof{ I^{ST}(u_{t}^\omega,u_{t-1}^\omega)}\\
\mathrm{s.t.} \quad & g_{t}^\omega+ V_{t}- \delta_{t}^\omega = b^\omega_t \quad \nonumber \\
&   u_{t}^\omega \underline{g}_{t} \le g_{t}^\omega \le u_{t}^\omega \Bar{g}_{t}\nonumber \\
&u_{t}  \in \{0,1\} 
\quad\quad\quad\quad \ftint , \mathrm{a.e.} \, \omega \in \Omega .
\end{align}
The function $I^{ST}(u_{t}^\omega,u_{t-1}^\omega)$ is non-linear and depends on both $\delta_t^\omega$ and $\delta_{t-1}^\omega$. 
In consequence, the joint predictive distributions $P(\delta_t^\omega , \delta_{t-1}^\omega)$ are required.
This distinction is useful, as there are more methods to predict a bivariate distribution $P(\delta_t^\omega , \delta_{t-1}^\omega)$ than a multivariate distribution $P(\vdelta^\omega)$.
\\  
\textbf{Required forecast:} Expected price, for adjacent time steps joint residual demand distributions.

\subsubsection{Uncertain price and residual demand, with storage}
In this setup, we consider scheduling the electricity sale or purchase for the provision of an uncertain residual demand with help of an energy storage. 
This can be expressed as
\begin{align}
\min_{\vs^\omega, \vb^\omega,\vV} \quad  & \sum_{t\in  \setT}    \EVof{\pi_{t}^\omega}  V_{t} + c^{b+}\EVof{ [b_t^\omega]_+ }+ c^{b-}\EVof{ [b_t^\omega]_-}\\ 
\mathrm{s.t.} \quad & V_{t} + s_{t-1}^\omega  - s_t^\omega - \delta_t^\omega = b_t^\omega  \nonumber \\
&   \underline{s}_{t} \le s_{t}^\omega \le \Bar{s}_{t} 
\quad \quad \quad \quad  \ftint , \mathrm{a.e.}\, \omega \in \Omega. 
\end{align}
The feasibility of $b_t^\omega$, which is non-linear in the expectation, is dependent on the realization of $\delta_t^\omega$. 
Due to the coupling between all adjacent time steps in the energy balance constraint, it is also depending on all other $\delta_{t\in \setT}^\omega$.
A joint predictive distribution is therefore necessary for all time steps of the residual demand.
\\
\textbf{Required forecast:} Expected price, full joint residual demand distribution.

\section{Bidding curve optimization}
\label{sec:bidding}
We now examine the setup, in which the operator can submit a bidding curve.
In this case the accepted volume $V_t(\pi^\omega_t)$ depends on the realized market price $\pi^\omega_t$.
\subsubsection{Uncertain price, fixed residual demand, power plant without startup costs}
First, we consider optimizing the bidding curve for selling the generation of a single power plant at an uncertain price without residual demand or startup costs.

For comprehensibility we first assume no demand. We can then substitute $g_t^\omega$ in the cost function with $V(\pi_t^\omega)$ and obtain
\begin{equation}  
    \label{eq:bid_nocoupling}
   \max_{ \vV(.) \in \mathcal{V}} \quad \sum_{t\in \setT}  \EVof{(\pi_t^\omega-c_t^F) V_t(\pi_t^\omega) }.
\end{equation}
Expressing the expectation as an integral split at the cost $c_t^F$ yields
\begin{align}
      \max_{ V(.) \in \mathcal{V}} \quad \sum_{t \in \setT} \int_{-\infty}^{c_t^F} (\omega-c_t^F)V_t(\omega) f_\pi(\omega) d\omega \nonumber \\
      +  \int_{c_t^F}^{\infty} (\omega-c_t^F)V_t(\omega) f_\pi(\omega) d\omega .
\end{align}
Let us examine the first integral. 
If $\omega < c_t^F$, then the term $\omega-c_t^F$ is always negative. 
Due to the property of a probability density function we know $f_{\pi_t^\omega}(\omega) > 0$, and thus it follows that an optimal bidding curve must have $V^*_t(\pi_t^\omega) = V^{min}$ for $\pi_t^\omega < c_t^F$, independently of $f_{\pi_t}$. 
Similarly, one can show that $V^*_t(\pi_t^\omega) = V^{max}$ for $\pi_t^\omega > c_t^F$, again independently of $f_{\pi_t}$.
Therefore, a prediction of the price is not necessary.
The values of $V^{min}$ and $V^{max}$ depend on generation limits and demand.
This setup applies for power plants with negligible startup costs as well as power plants that almost always run because of very high startup costs, e.g. lignite power plants. 
It can also be generalized to the case of multiple power plants, for which the integral has to be split into multiple intervals in ascending order given by the generation cost of each plant.
\\
\textbf{Required forecast:} No forecast required.
\subsubsection{Uncertain price, fixed residual demand, power plant with startup costs}
\label{subsec:bid_ppSC}
We again consider a single power plant, but now include startup costs in the setup. 
The formulation then is
\begin{align}
\min_{\vV(.),\vg^\omega, \vu^\omega }  &\quad \sum_{t\in \setT}  \EVof{\pi_t^\omega V_t(\pi_t^\omega)} + c_t^F \EVof{g_t^\omega}  \nonumber \\
\label{eq:bis_ppSC_cf}
&\quad+  c^{ST} \EVof{ I^{ST}(u_{t}^\omega,u_{t-1}^\omega)}\\ 
\mathrm{s.t.} \quad& g_{t}^\omega+ V_t(\pi_t^\omega)- d_t = 0 \quad \nonumber \\
&   u_{t}^\omega \underline{g}_{t} \le g_{t}^\omega \le u_{t}^\omega \Bar{g}_{t} \nonumber \\
&u_{t}  \in \{0,1\},V_t(.) \in \mathcal{V}
\quad  \ftint , \mathrm{a.e.} \, \omega \in \Omega .
\label{eq:bis_ppSC_st}
\end{align}  
The cost function includes the expectation of a non-linear combination of $u_t^\omega$ and $u_{t-1}^\omega$, whose feasibility is linked to $\pi_t^\omega$ and $\pi_{t-1}^\omega$, respectively. 
It follows that pairwise joint predictive distributions $P(\pi_t^\omega,\pi_{t-1}^\omega)$ are necessary.
\\
\textbf{Required forecast:} Pairwise joint price distributions.
\subsubsection{Uncertain price and residual demand, procurement only}
Considering imbalance cost and an uncertain residual demand, we obtain 
\begin{align}
\label{eq:bid_uncdemand}
   \min_{\vb^\omega,\vV(.) }\quad & \sum_{t\in \setT}   \EVof{\pi_t^\omega V_t(\pi_t^\omega)} \nonumber \\
   & \; + c^{b+} \EVof{[b_t^\omega]_+ }+ c^{b-} \EVof{ [b_t^\omega]_-} \\
   \mathrm{s.t.}\quad & V_t(\pi_t^\omega) - \delta_t^\omega = b_t^\omega  \quad \nonumber \\
   &V_t(.) \in \mathcal{V}\quad \ftint , \mathrm{a.e.}\, \omega \in \Omega .
\end{align}  
The cost function contains the expectation of the positive and negative functions, which are non-linear with respect to both  $\pi_t^\omega$ and $\delta_t^\omega$.
Thus, the joint distribution $P(\delta_t^\omega,\pi_t^\omega)$ is necessary.   
\\
\textbf{Required forecast:} Joint price and residual demand distribution for each time step.
\subsubsection{Uncertain price and residual demand, with startup costs}
We again consider the setup with a power plant, but consider the residual demand as uncertain.
\begin{align}
\min_{\vV(.),\vb^\omega,\vg^\omega,\vu^\omega }& \sum_{t\in \setT}  \EVof{\pi_t^\omega V_t(\pi_t^\omega)} +c_t^F \EVof{g_t^\omega} \nonumber\\
&\quad + c^{b+} \EVof{[b_t^\omega]_+ }+ c^{b-} \EVof{ [b_t^\omega]_-} \nonumber \\
&\quad+  c^{ST} \EVof{ I^{ST}(u_{t}^\omega,u_{t-1}^\omega)}\\ 
\mathrm{s.t.}\quad& g_{t}^\omega+V_t(\pi_t^\omega) - \delta_t^\omega = b_t^\omega  \quad \nonumber \\
&   u_{t}^\omega \underline{g}_{t} \le g_{t}^\omega \le u_{t}^\omega \Bar{g}_{t} \nonumber \\
&u_{t}  \in \{0,1\}, V_t(.) \in \mathcal{V}
\quad  \ftint , \mathrm{a.e.} \, \omega \in \Omega
\end{align}  
One of the constraints on $u_t^\omega$ can be formulated as $u_t^\omega \le \underline{g}_{t}( b_t^\omega - V_t(\pi_t^\omega) + \delta_t^\omega)$, i.e., the feasibility of $u_t^\omega$ depends on $\pi_t^\omega$ and $\delta_t^\omega$.
Hence, the non-linear function $I^{ST}(u_{t}^\omega,u_{t-1}^\omega)$ depends on  $\pi_t^\omega$,$\pi_{t-1}^\omega$,$\delta_t^\omega$, and $\delta_{t-1}^\omega$.
Thus, the predictive joint distributions $P(\pi_t^\omega,\pi_{t-1}^\omega,\delta_t^\omega,\delta_{t-1}^\omega)$ are required.
\\
\textbf{Required forecast:} Price and residual demand distributions joint for adjacent time steps.

\subsubsection{Uncertain price and residual demand, with storage}
Let us now consider a setup with an energy storage, in which both price and residual demand are uncertain. The optimization problem then is
\begin{align}
\min_{\vV(.),\vb^\omega,\vs^\omega }& \sum_{t\in \setT}  \EVof{\pi_t^\omega V_t(\pi_t^\omega)}+ c^{b+} \EVof{[b_t^\omega]_+ }+ c^{b-} \EVof{ [b_t^\omega]_-} \\
\mathrm{s.t.}\quad & V_t(\pi_t^\omega) + s_{t-1}^\omega  - s_t^\omega - \delta_t^\omega = b_t^\omega  \quad \nonumber \\
&   \underline{s}_{t} \le s_{t}^\omega \le \Bar{s}_{t}  \nonumber \\
&V_t(.) \in \mathcal{V}
\quad\quad\quad\quad\quad  \ftint , \mathrm{a.e.} \, \omega \in \Omega .
\end{align}  
The balance equation's feasibility is dependent on both random variables and is, due to the storage balance, coupled between all time steps. 
Thus, the distribution $P(\vpi^\omega,\vdelta^\omega)$ is required.
\\
\textbf{Required forecast:} Full joint price and residual demand distribution.
\section{Experiment: Bidding with Variable Time-Step Coupling}
\label{sec:experiment}

\begin{figure}[b]
	\centering
	\includegraphics[width=1\linewidth]{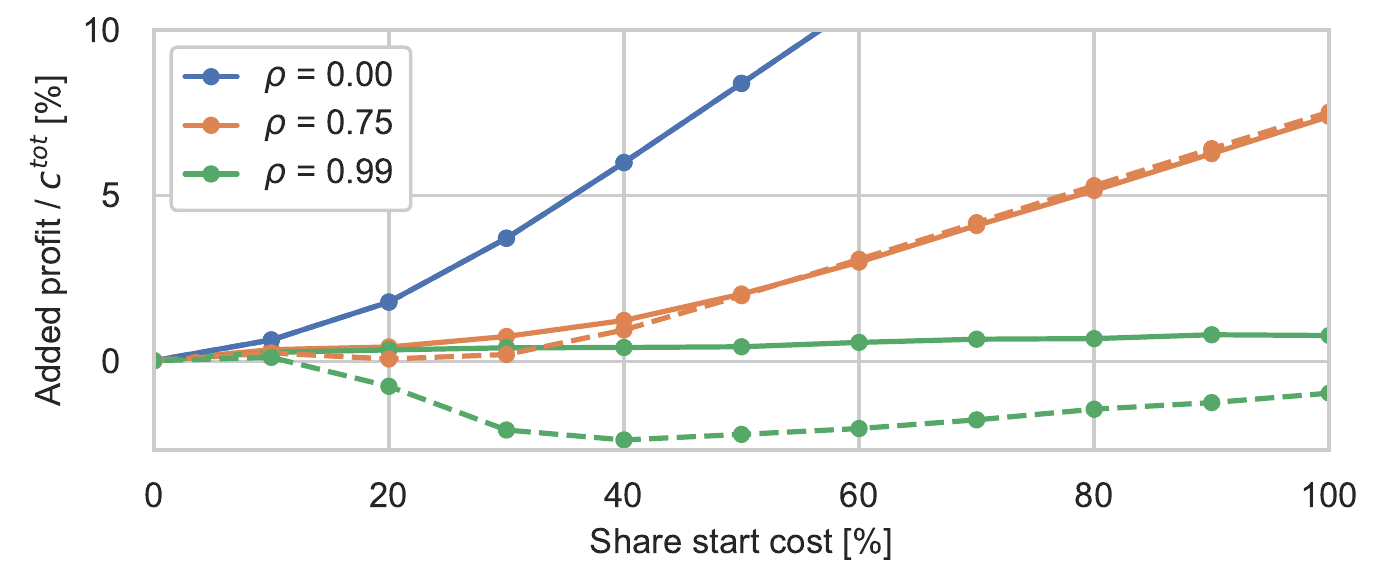}
	\caption{Expected additional profit when employing a multivariate forecast (solid) or a marginal forecast (dashed) compared to the expected value model as a function of the share of startup costs in $c^{tot}$. The line colors correspond to values of the true price correlation parameter $\rho$. The added profit is shown as a percentage of $c^{tot}$. All results are averages of 10 runs.}
	\label{fig:bid_stVed}
\end{figure}
\begin{table*}[h!]
    \caption{Required forecasts for each setup.}
    \centering
	\includegraphics[width=0.98\linewidth]{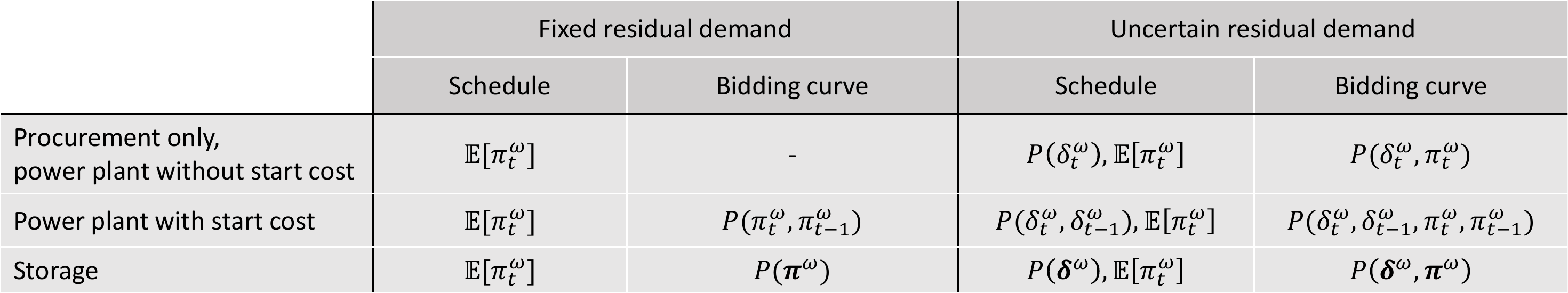}
    \label{tab:disc_summary}
\end{table*}


So far, we analyzed theoretically which type of probabilistic forecast can improve the solution of a stochastic program. 
In the following simulation experiment we evaluate this for a concrete example. 
We moreover demonstrate that the transition between setups that require different forecast types can actually be gradual. 

\subsection{Setup}
We evaluate the bidding curve optimization problem for a power plant without residual demand but with startup costs and uncertain market prices.
The importance of implied time step coupling is varied by changing the share of the startup costs in the total cost. 
Following our outline above, this should affect the advantage of using a probabilistic forecast of the pairwise price distribution compared to an expected value forecast.

For our implementation we apply the sample average approximation and replace the expectation in  (\ref{eq:bis_ppSC_cf}) and (\ref{eq:bis_ppSC_st}) with an empirical mean over a set of scenarios.
The optimization problem then reads 
\begin{align}
\min_{\vV(.),\vg,\vu} \quad & \sum_{s\in \setS} w_s \sum_{t\in \setT}  (\pi_{t,s} - c^F) V_t(\pi_{t,s}) \nonumber \\
&\quad+  c^{ST}I^{ST}(u_{t,s},u_{t-1,s})\\ 
\mathrm{s.t.} \quad& g_{t,s}+ V_t(\pi_{t,s}) = 0 \quad \nonumber \\
&   u_{t,s} \underline{g}_{t} \le g_{t,s} \le u_{t,s} \Bar{g}_{t} \nonumber \\
&u_{t,s}  \in \{0,1\},V_t(.) \in \mathcal{V}
\quad\quad  \ftint, \forall s \in \setS ,
\end{align}  
with $\setS$ being the set of all scenarios and $w_s$ the corresponding weights.
The bidding curves $V_t(\pi_{t,s})$ are modeled as an increasing step function. 
The optimization horizon consists only of two time steps, with the power plant being in off state initially.

In this experiment, we consider the true price distribution to be a multivariate Gaussian, i.e., $P(\vpi^\omega) = \mathcal{N}(\mathbf{\vmu},\Sigma)$. The covariance matrix $\Sigma$ is constructed from the variance $\sigma^2$ and the correlation parameter $\rho$.
We use 300 equally weighted scenarios generated from the forecast distributions.
We add 30 uniformly distributed bid scenarios to allow the optimization model to choose bid prices outside of the sampled scenarios.
To reduce their influence on the forecast, the set of bid scenarios has the same mean as the forecast scenarios and is weighted with, in sum, $1 \%$ of the total weight.
We consider three different types of forecasts: the multivariate model which predicts the correct joint distribution $\mathcal{N}(\mathbf{\vmu},\Sigma)$, the marginal model which predicts $\mathcal{N}(\mathbf{\vmu},\sigma \mathbf{I})$, and the expected value model which only predicts $\mathbf{\vmu}$.
To ensure comparability, we keep the total possible cost $c^{tot} = c^{ST}+2 c^{F}$ constant and vary the share of startup costs in the total cost $\frac{c^{ST}}{c^{tot}}$.


Note that the true distribution influences the relevance of the type of forecast as well, e.g., if the probability of $\sum_{t\in \setT}\pi_t^\omega > c^{tot}$ is extremely low, probabilistic and point forecasts will perform equally well.
Therefore, we chose $\vmu=(0.45,0.45)^T c^{tot}$ and $\sigma=0.1 c^{tot}$ such that the resulting schedule is not obvious.

We solve the 2-stage stochastic program with all three forecast types, obtain each of their solutions, i.e., the optimal bidding curves, and then calculate the expected profit in the second stage model for each of solution.
The expected profit is estimated with 10000 realizations of the true distribution.
In the following, we analyze the added profit of using the marginal distribution over the expected value and the added profit of using the multivariate distribution over the expected value.
The model is implemented with CVXPY \cite{diamond2016_cvxpy} and solved with Gurobi \cite{gurobi}. 

\subsection{Results}
Figure \ref{fig:bid_stVed} confirms our theoretical findings that only if $c^{ST}>0$, a probabilistic (marginal or multivariate) forecast can offer a benefit over the expected value forecast. 
However, even though a distribution is theoretically required to achieve the best possible results, the benefits of a probabilistic model differ depending on the parameters of the optimized setup:
For highly flexible setups, i.e., a low share of startup costs, the added profit from using the probabilistic forecast is small but increases with higher shares of startup costs.
Moreover, the added profit of using a probabilistic forecast is influenced by the correlation of the true distribution, with a higher correlation resulting in a lower advantage.
Comparing the results of marginal and multivariate model shows that the advantage of the multivariate model is only significant for very strong correlations.


\section{Discussion}
\label{sec:discuss}
We theoretically analyzed the stochastic programming formulations of scheduling and bidding in energy markets under uncertainty with respect to the required type of forecast.

We found that in the considered market schedule optimization setups a probabilistic price forecast is not necessary, whereas a probabilistic demand and renewable generation forecast mostly is.
Bidding curve optimization requires forecast distributions, joint with respect to the price and the residual demand. 
For both market schedule and bidding curve optimization, setups including an energy storage require a full temporal joint distribution and those containing power plants with startup costs require distributions joint between adjacent time steps. 
Without any time step coupling, distributions that are marginal regarding the time steps are sufficient.
Table \ref{tab:disc_summary} summarizes these findings.


Future work could extend the analysis to other types of optimization problems, such as optimal power flow, or analyze the quantitative value of forecasts in more depth.

\bibliographystyle{IEEEtran}
\bibliography{beykirch_pmaps.bib}

\begin{thebibliography}{10}
\providecommand{\url}[1]{#1}
\csname url@samestyle\endcsname
\providecommand{\newblock}{\relax}
\providecommand{\bibinfo}[2]{#2}
\providecommand{\BIBentrySTDinterwordspacing}{\spaceskip=0pt\relax}
\providecommand{\BIBentryALTinterwordstretchfactor}{4}
\providecommand{\BIBentryALTinterwordspacing}{\spaceskip=\fontdimen2\font plus
\BIBentryALTinterwordstretchfactor\fontdimen3\font minus
  \fontdimen4\font\relax}
\providecommand{\BIBforeignlanguage}[2]{{%
\expandafter\ifx\csname l@#1\endcsname\relax
\typeout{** WARNING: IEEEtran.bst: No hyphenation pattern has been}%
\typeout{** loaded for the language `#1'. Using the pattern for}%
\typeout{** the default language instead.}%
\else
\language=\csname l@#1\endcsname
\fi
#2}}
\providecommand{\BIBdecl}{\relax}
\BIBdecl

\bibitem{Maggioni2012_analyzingEvSolution}
F.~Maggioni and S.~W. Wallace, ``Analyzing the quality of the expected value
  solution in stochastic programming,'' \emph{Annals of Operations Research},
  vol. 200, no.~1, pp. 37--54, 2012.

\bibitem{kumbartzky2017_OptimalChpOperation}
N.~Kumbartzky, M.~Schacht, K.~Schulz, and B.~Werners, ``Optimal operation of a
  chp plant participating in the german electricity balancing and day-ahead
  spot market,'' \emph{European Journal of Operational Research}, vol. 261,
  no.~1, pp. 390--404, 2017.

\bibitem{Ottesen2016_prosumerBidding}
S.~{\O}. Ottesen, A.~Tomasgard, and S.-E. Fleten, ``Prosumer bidding and
  scheduling in electricity markets,'' \emph{Energy}, vol.~94, pp. 828--843,
  2016.

\bibitem{Ackermann2019}
S.~Ackermann, A.~Szabo, S.~Paulus, and F.~Steinke, ``Comparison of two
  day-ahead offering strategies for a flexible chp plant in germany,'' in
  \emph{2019 IEEE PES Innovative Smart Grid Technologies Europe (ISGT-Europe)},
  2019.

\bibitem{gneiting2014_probabilisticFc}
T.~Gneiting and M.~Katzfuss, ``Probabilistic forecasting,'' \emph{Annual Review
  of Statistics and Its Application}, vol.~1, pp. 125--151, 2014.

\bibitem{dumas2022_deepGenModProbFc}
J.~Dumas, A.~Wehenkel, D.~Lanaspeze, B.~Corn{\'e}lusse, and A.~Sutera, ``A deep
  generative model for probabilistic energy forecasting in power systems:
  normalizing flows,'' \emph{Applied Energy}, vol. 305, p. 117871, 2022.

\bibitem{toubeau2018_deepLearningProbFc}
J.-F. Toubeau, J.~Bottieau, F.~Vall{\'e}e, and Z.~De~Gr{\`e}ve, ``Deep
  learning-based multivariate probabilistic forecasting for short-term
  scheduling in power markets,'' \emph{IEEE Transactions on Power Systems},
  vol.~34, no.~2, pp. 1203--1215, 2018.

\bibitem{janke2020_priceFcImplGenMods}
T.~Janke and F.~Steinke, ``Probabilistic multivariate electricity price
  forecasting using implicit generative ensemble post-processing,'' in
  \emph{2020 International Conference on Probabilistic Methods Applied to Power
  Systems (PMAPS)}.\hskip 1em plus 0.5em minus 0.4em\relax IEEE, 2020, pp.
  1--6.

\bibitem{Hong2020_EnergyForecastingOutlook}
T.~Hong, P.~Pinson, Y.~Wang, R.~Weron, D.~Yang, and H.~Zareipour, ``Energy
  forecasting: A review and outlook,'' \emph{IEEE Open Access Journal of Power
  and Energy}, 2020.

\bibitem{Shapiro2021_lecturesOnStochProg}
A.~Shapiro, D.~Dentcheva, and A.~Ruszczynski, \emph{Lectures on stochastic
  programming: modeling and theory}.\hskip 1em plus 0.5em minus 0.4em\relax
  SIAM, 2021.

\bibitem{diamond2016_cvxpy}
S.~Diamond and S.~Boyd, ``{CVXPY}: {A} {P}ython-embedded modeling language for
  convex optimization,'' \emph{Journal of Machine Learning Research}, vol.~17,
  no.~83, pp. 1--5, 2016.

\bibitem{gurobi}
\BIBentryALTinterwordspacing
{Gurobi Optimization, LLC}, ``{Gurobi Optimizer Reference Manual},'' 2021.
  [Online]. Available: \url{https://www.gurobi.com}
\BIBentrySTDinterwordspacing

\end{thebibliography}

\end{document}